# Survey of Vacuum UltraViolet experimental data in relation to radiation characterization for Earth high-speed re-entry

Philippe Reynier[1]

*Ingénierie et Systèmes Avancés, Cestas, France*

This contribution is a survey of the available experimental radiation data measured in the VUV range related to hypersonic atmospheric entry. The objective is to identify the experimental datasets already gathered during aerothermodynamics studies for preparing sample return missions, and future. The final goal is to identify the most valuable VUV datasets for comparisons with future measurements to be performed in the European shock-tube ESTHER. Due to the limited number of studies covering VUV radiation in relation to space exploration missions, and manned Moon exploration, the review has been extended to domains such as nuclear fusion, exobiology, chemical and process engineering.

## Nomenclature

**Acronyms:**

CFD:    Computational Fluid Dynamics;
CNRS:   Centre National de la Recherche Scientifique
EAST:   Electric Arc Shock Tube;
ESA:    European Space Agency;

---

[1] Research Engineer, ISA, 16 chemin de l'Ousteau de Haut, 33610 Cestas, France, Philippe.Reynier@isa-space.eu

| | |
|---|---|
| HVST: | High Velocity Shock Tube |
| ICP: | Induced Coupled Plasma |
| IR: | InfraRed; |
| IRS: | Institute für RaumfahrtSysteme |
| IST: | Instituto Superior Tecnico |
| JAXA: | Japan Aerospace eXploration Agency; |
| LENS: | Large Energy National Shock-tunnel; |
| LIF: | Laser Induced Fluorescence; |
| NASA: | National Aeronautics and Space Administration; |
| NIR: | Near InfraRed; |
| TPS: | Thermal Protection System; |
| UV: | Ultra Violet; |
| VIS: | VISible; |
| VUV: | Vacuum Ultra Violet. |
| XUV : | EeXtreme Ultra Violet (from 10 to 124 nm). |

**1 Introduction**

The European Space Agency (ESA) is actually supporting the development of a high-velocity shock-tube, ESTHER [1-2] for preparing future exploration missions of the solar system. These missions could potentially be manned missions to the Moon, or sample return missions to comets and asteroids involving the superorbital Earth re-entry of a return capsule. Such entries are characterized by high heat-fluxes and the return capsule has to endure severe entry conditions in terms of radiative and convective heating. Such high levels of heat-fluxes characterize all sample return missions performed so far, like Stardust [3], Genesis [4], and Hayabusa [5].

The current effort focuses on the support of the European shock tube ESTHER [6], and its instrumentation. ESTHER is a facility under development by an international consortium led by IST of Lisbon, under funding from the European Space Agency. It is a two-stage combustion driven shock-tube, as shown in Figure 1, with laser ignition. ESTHER has a 16 m length, a main section of 80 mm diameter, and can reach shock velocities from 4 up to 12 km/s for Earth atmospheric entry, with pre-shock pressures in between 0.1 and 100 mbar (10 to 10 000 Pa). ESTHER's capabilities are going to be extended via the integration of a system measuring vacuum ultraviolet (VUV) radiation. Such instrumentation will be of interest for high-speed Earth, Venus entries, and also Mars entries with radiation from CO4+ system. The goal of the current survey is to support the development of this VUV instrumentation by identifying the most relevant experimental datasets for air in the perspective of potential cross-checking with other facilities.

Measurements in VUV range cover wavelengths from 62 to 200 nm (6-20 eV). In hypersonic flows, radiative heating can be significant in this range of the spectrum; this particularly applies to sample return mission. Unfortunately, VUV emission is difficult to investigate due to the instrumentation limitations: in ground tests VUV radiation is absorbed by the ambient oxygen. Usually dedicated deuterium lamps [7] are used for such measurements. These devices required a specific calibration based on advanced photon metrology [8].

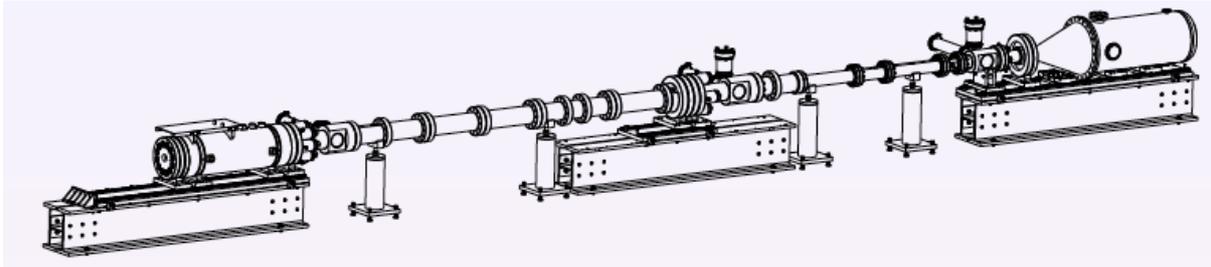

**Figure 1: Sketch of ESTHER shock-tube (credit IST)**

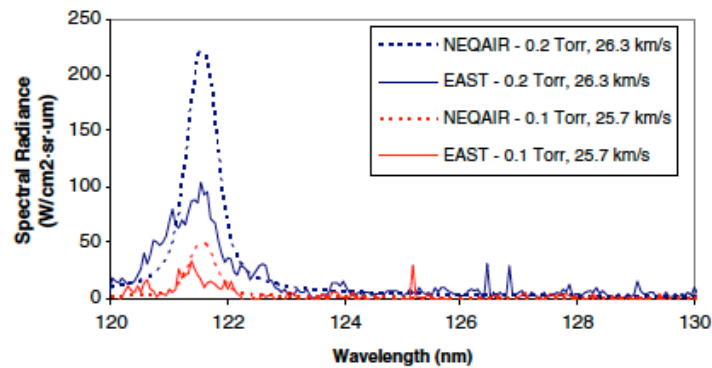

**Figure 2: VUV spectra (1 Torr = 133 Pa) and NEQAIR reconstruction at 26 km/s (from[10])**

Due to the limited amount of studies performed so far focusing on VUV radiation in relation with hypersonic entry, the focus has been extended to other domains such as tokamaks, chemistry applications, and exobiology. It has to be noted that VUV measurements have been also performed for Titan [9], Uranus, and Saturn [10] entry conditions. In both cases, these measurements were performed in the NASA EAST shock-tube. There were done for wavelength down to 115 nm for Titan aerocapture conditions at pressures of 13.3 and 133 Pa, and velocities ranging in 5-9 km/s. However most of the effort was focused on CN radiation and very little is available on the results obtained for the VUV radiation range. More material is available in [10] where Cruden & Bogdanoff have assessed radiation for Saturn and Uranus entries. Tests were performed for a mixture of $H_2$-He (89 and 11 % in volume fractions), pressure of 13 and 66 Pa, and velocities ranging from 20 to 30 km/s. The results showed that radiation is significant for velocities higher than 25 km/s. Radiation was investigated from VUV to IR, for atomic hydrogen Lyman, Balmer, and Palmer lines, and $H_2$ Lyman band. VUV radiation corresponds to H Lyman $\alpha$ line and $H_2$ Lyman band. Figure 2 shows some experimental results for the VUV band compared to those provide by the radiation tool NEQAIR [11]. For the current effort, the emphasis has been put on Earth superorbital re-entry conditions, the paper has been focused on VUV data only, as a consequence, other existing data for other parts of the spectrum obtained in many facilities have not been considered, since they are outside the scope of this survey.

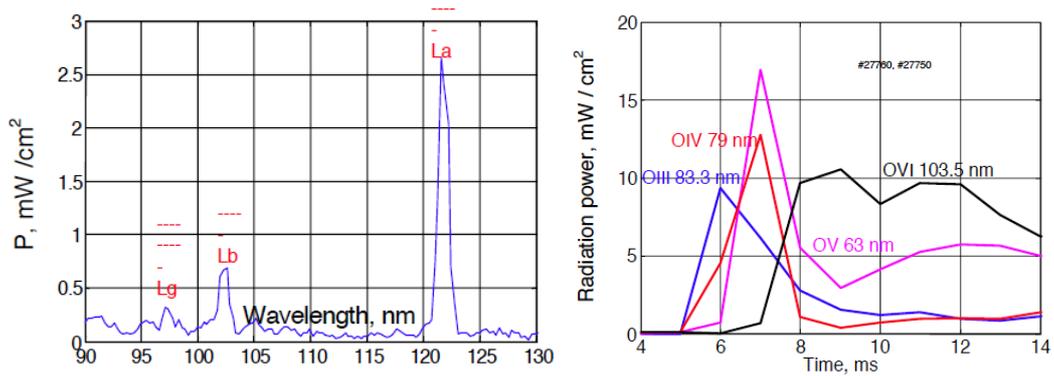

**Figure 3:** Spectrum of plasma radiation discharge during the first ms in a tokamak, with hydrogen lines (from [12])

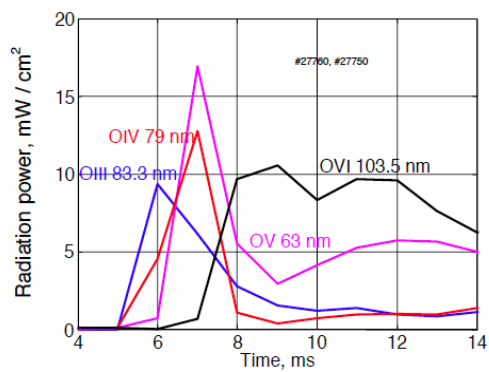

**Figure 4:** Time history of line radiation emission for different ionised oxygen lines during the CASTOR plasma discharge (from [12])

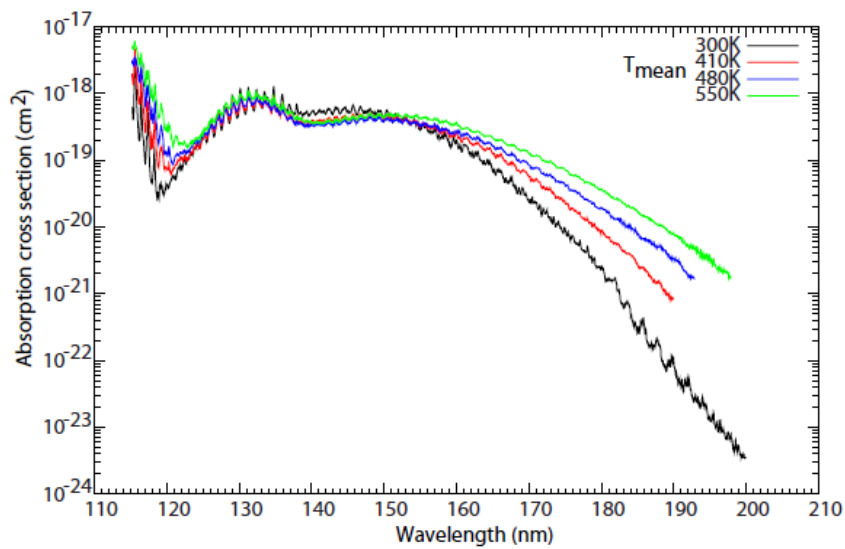

**Figure 5:** Absorption cross section of $CO_2$ for wavelengths between 115 and 220 nm, at different temperatures: 300 K (black, 410 K (red), 480 K (green), and 550 K (blue) (from [14])

## 1 Non-aerospace data

VUV measurements are also carried out for nuclear, exobiology, chemistry, and process engineering applications. In this section, some elements found in the literature are given. They might be of interest for future space missions or to extended ESTHER use to other applications, or to adapt innovative measurement techniques to shock-tube testing.

### 1.1 Nuclear fusion

VUV test campaigns have been undertaken to characterize the hydrogen plasma in order to estimate the level of impurity content [12], radiation losses, and the full-radiated power. For that objective, VUV measurements in tokamak have been already performed: for example Piffl et al [12] have developed this technique using a tungsten lamp and have performed radiation measurements in a tokamak for a 90-130 nm wavelength range. Their objective was to estimate the radiation emitted in the hydrogen Lβ (102.6 nm) and Hα lines (656.3 nm). A spectrum covering the first miliseconds of discharge when hydrogen lines dominate the spectrum is shown in Figure 3. These lines are used for calibration, which is not straightforward, since hydrogen Lβ line lies near the bright oxygen doublet OVI (103.2, 103.7 nm). Impurities are due to oxygen ionization and dedicated measurements are performed in the VUV range for their estimate as highlighted in Figure 4.

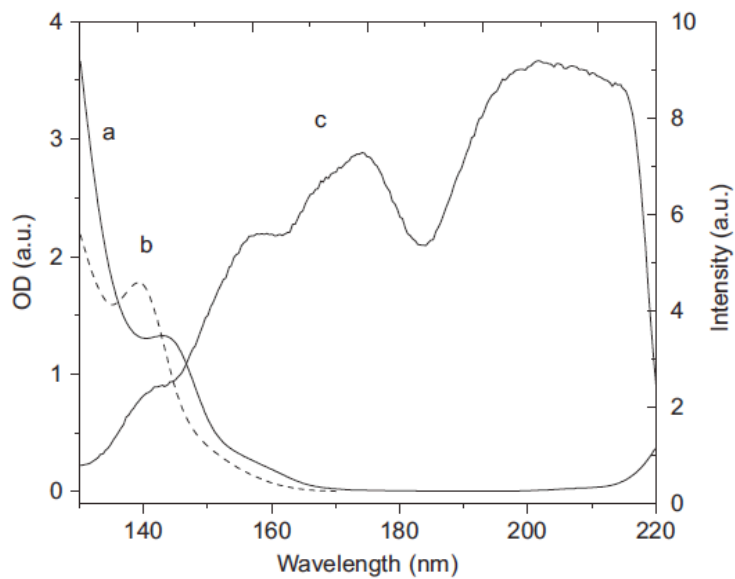

**Figure 6: a) Absorption for undoped KYF$_4$; b) Thermoluminescence excitation; c) Photoexcitation (255 nm) spectra in the VUV region for KYF$_4$: Pr$^{3+}$ (from [18])**

Another interest for the application of the VUV spectrometry in tokamaks, is related to the plasma/surface interaction [13]. Vasenin et al [13] have performed VUV measurements of plasma-surface interaction with a supersonic plasma of hydrogen injected with a gun over a wavelength range from 0 up to 40 nm. The jet was impinging on models made of graphite or tungsten. Measurements highlight that VUV radiation plays an important part in energy balance of plasma/surface interaction. VUV emission duration proves to be noticeably less than plasma stream duration. Such data could be of interest for future exploration missions to the giant planets since the interaction between the plasma flow of helium and hydrogen and the carbon phenolic is a key issue when sizing the thermal protection system (TPS).

## 1.2 Exobiology

UV and VUV absorption is an essential ingredient of photochemical atmosphere planets. Exoplanet research has discovered a large population of hot Jupiter and hot Neptune (giant planets orbiting close to their stars). These exoplanets with temperature in between 400 and 2500 K can be studied using spectroscopy. However, at these temperatures UV photolysis cross-section data are lacking. As a consequence, investigations [14-15] to provide missing data such as cross-sections for exoplanet atmosphere components and high temperatures atmospheres, are currently carried out for closing this gap. An important atmospheric component for some of these exoplanets is $CO_2$ that is observed and photodissociated in such atmospheres. For obtaining cross sections for this molecule (see Figure 5), Vénot et al [14] have performed VUV-absorption measurements of $CO_2$ using synchrotron radiation. In the 115-200 nm range a tunable VUV light source was used, and a deuterium lamp in the 195-230 nm range. Measurements have been conducted for temperatures up to 800 K, they demonstrated that the cross section of $CO_2$ is very sensitive to temperature especially above 160 nm.

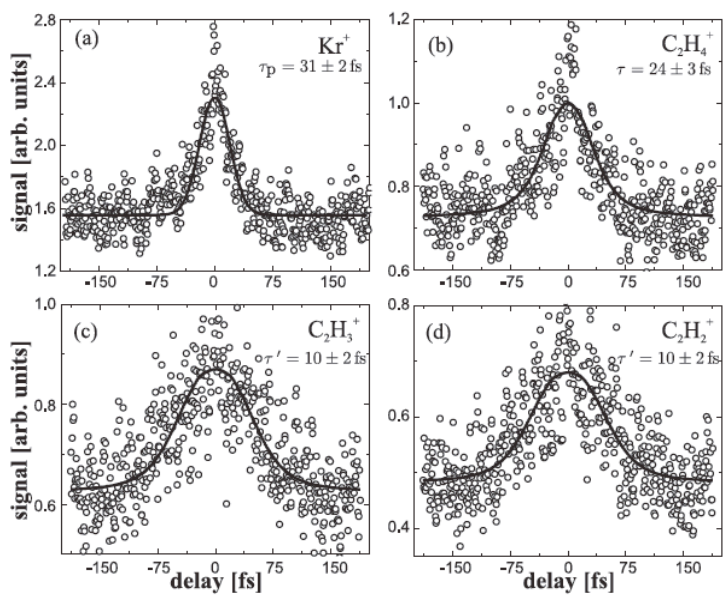

**Figure 7: Time resolved two-photon ionization of krypton and ethylene: Solid lines – numerical simulations (from [21])**

Other studies [15-16] have been conducted for investigating VUV photodissociation of prebiotic molecules that are precursors to life, like small prebiotic molecules such as amino acids, purin, pyrimidines including nucleobases. VUV radiation is among the important energy sources impinging on these molecules in astrophysical sites. A recent study [16] has been focused on acetamide ($H_3C-C(O)-NH_2$) which is one simple model molecules for the peptide linkage in polypeptides and proteins. The VUV photoionization of this molecule has been studied using synchrotron radiation for a 8-24 eV energy range.

## 1.3 Chemistry and Process Engineering

Study of VUV irradiation is of interest for several applications related to chemistry and process engineering. VUV irradiation is used for the conversion of natural gas has highlighted in the recent study of Orlovskii et al [17]. This experimental investigation demonstrated that VUV irradiation at a wavelength of 172 nm has for effect to strongly influence phase transitions in natural gas. A decrease of water concentration by a factor 11 and an increase of condensate extraction by a factor in between 2 and 16 were observed. This is due to water photolysis forming high reactive OH and H radicals that react with hydrocarbons.

The same technique can be used for doping fluoride crystals [18]. Some of them such as $CaF_2$ are transparent in a wide spectral range of the spectrum, from VUV to IR, and are therefore widely used as optical materials. Kristianpoller et al [18] have investigated the irradiation effects and dosimetric properties for some of these crystals ($CaF_2$, $KYF_4$ and $CsY_2F_7$) doped with $Pr^{3+}$. In particular, thermoluminescence excitation spectra have been measured in the 120-200 nm region, as shown in Figure 6. This provides the thermoluminescence sensitivity of the different crystals.

Plasma VUV radiation can also be of interest for characterizing dielectric thin films [19]. For that objective, the thin film is exposed to plasma VUV radiation in a cyclotron. Cismaru & Shohet [19] have measured the electrical conductivity of $SiO_2$ layers during exposure to argon and oxygen plasmas with controlled VUV emission. They have also modelled the VUV-induced conductivity of $SiO_2$. Their results show the interest of their measurements for the plasma processing of semiconductor devices. Other investigations were made to study the behaviour of thermal protection system materials to VUV radiation. This is particularly of interest for solar probes since there are submitted to an extreme environment characterized by high temperatures, particle bombardment, and VUV irradiation. Such conditions can be reproduced in solar furnaces allowing sample heating up to 2500 K, and high VUV fluxes for the H Lyman α line. Test campaigns [20] were conducted at CNRS Odeillo (France) for evaluating different composite materials.

Time resolved two VUV photon ionization measurements are of interest for molecular dynamics to study ultrafast phenomena [21]. Conde et al [21] have performed measurements in the VUV/XUV spectral region for investigating successfully the dynamics of excited ethylene and oxygen molecules as highlighted in Figure 7.

| Pressure (Pa) | Velocity (km/s) | Reference |
|---|---|---|
| 13.3, 26.6 | 9.5 to 15.5 | 25 |
| 13.3, 26.6, 40, 93.1, 133 | 10 | 27 |
| 13.3, 40, 133 | 10 | 26 |

**Table 1: VUV data obtained with EAST for Earth superorbital entry conditions**

## 2 Earth superorbital re-entry

Several test campaigns have been carried out focusing on VUV measurements for Earth re-entry conditions, this, in different facilities: NASA EAST shock-tube, X2 expansion tunnel at University of Queensland, plasma wind-tunnels at IRS and CNRS Orléans, ICP plasma torch at ECP ("Ecole Centrale de Paris"), and HVST JAXA shock-tube, where pure nitrogen tests [22] were performed to support the preparation of the Hayabusa mission. More details on most of these different facilities can be found in [2,23].

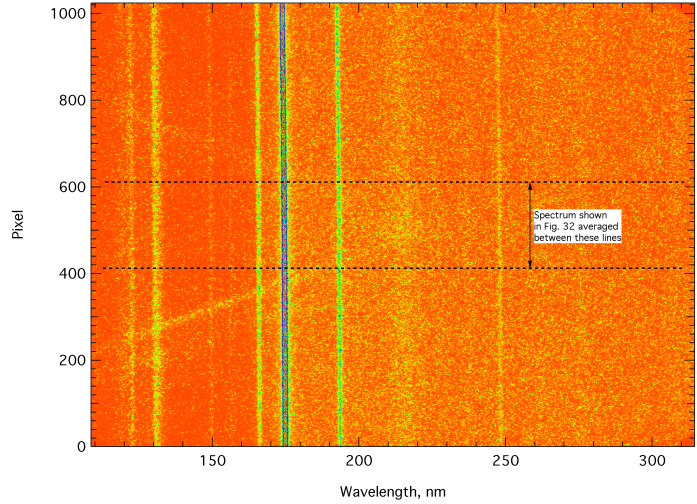

**Figure 8: VUV spectrum obtained in EAST (10.546 km/s, 266 Pa) (from [27])**

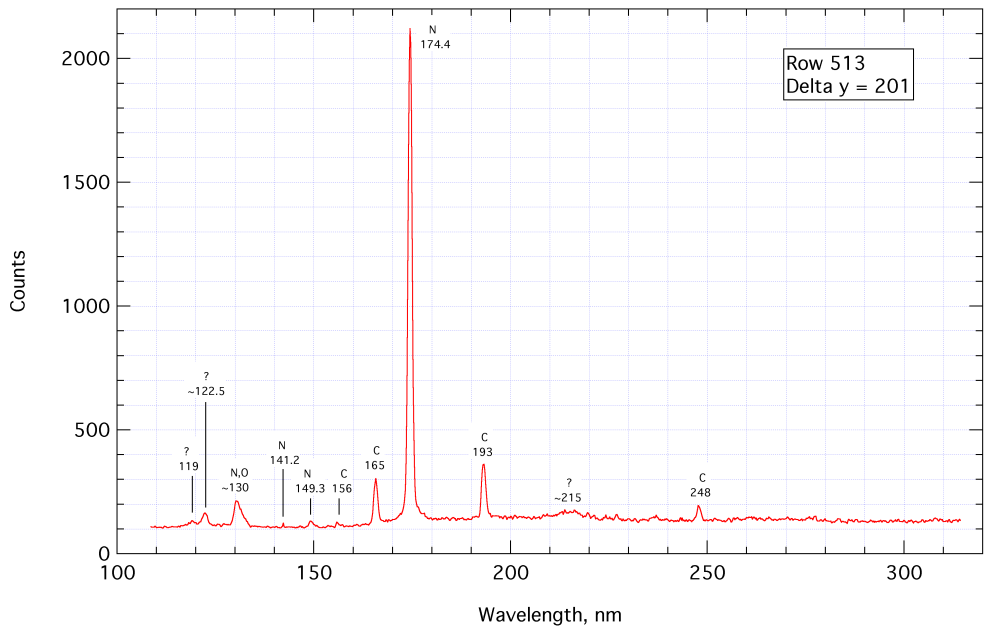

**Figure 9: VUV spectrum obtained in EAST (10.546 km/s, 26.6 Pa) with N, O and C lines (from [27])**

## 2.1 Shock facilities

Most of the experimental data covering the VUV range has been obtained in the EAST shock-tube, for supporting the Orion project [24] involving a Lunar return, and sample return missions [25]. In [26], if a test campaign involving VUV measurements in LENS XX, performed for the Orion project is mentioned, no spectrum covering the VUV range is provided. As a consequence, all data found in the literature for NASA projects on this topic have been obtained with EAST. The instrumentation available with EAST is detailed in [26-27], the calibration in [27]: it has been applied to Mars, Earth and Neptune entry conditions (shock velocity up to 34 km/s for Neptune). VUV, as well as UV, visible, and NIR can be investigated in this facility for a wavelength range from 110 up to 4000 nm. A typical spectrum obtained with the VUV spectrometer is shown in Figure 8. The tests performed in EAST for investigating Lunar return, and sample return conditions are resumed in Table 1.

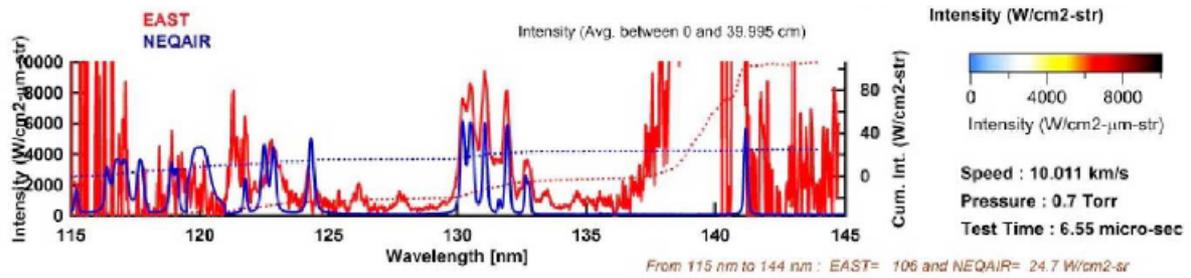

**Figure 10: Comparisons of EAST data and NEQUAIR results at 10 km/s and 93.1 Pa (from [28])**

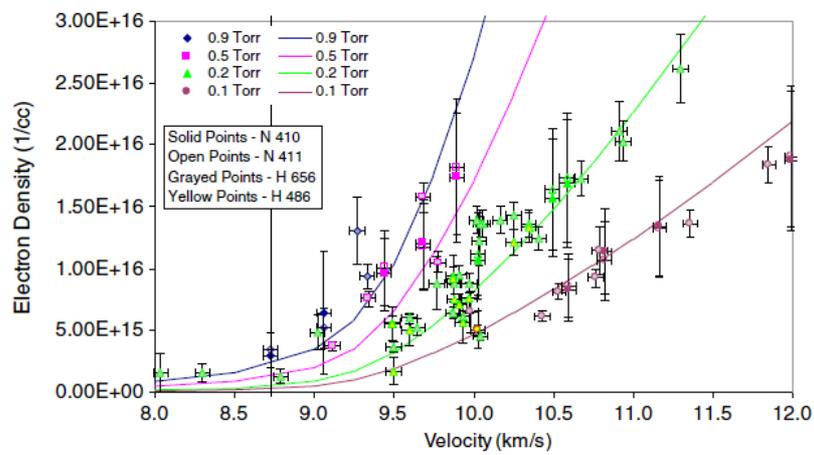

**Figure 11: Measurements of electron density for Lunar return conditions in EAST (from [29])**

Different spectra are available in [27], the VUV measurements are available for pressures of 26.6 and 93.1 Pa and velocities around 10 km/s. One spectrum with some O, N, and C lines obtained at a pressure of 26.6 Pa is shown in Figure 9. In a recent effort [25] the range of shock velocity from 9.5 up to 15.5 km/s has been extensively explored for 13.3 and 26.6 Pa.

Some results for the Orion project with the spectra obtained in EAST for the VUV range at 26.6, 40 and 93.1 Pa (around 10 km/s) have been published by Bose et al [28] and numerically reconstructed with NEQAIR [11]. An example of the results available for 93.1 Pa and 10 km/s is shown in Figure 10. These different test campaigns carried out in EAST have also allowed the measurements [29] of the electron density for Lunar return conditions as highlighted in Figure 11.

Experimental data for the VUV range have been detailed in [25] with the complete spectra obtained at 15.5 km/s (13.3 Torr) and 13.64 km/s (26.6 Torr). The experimental data has been numerically reconstructed with the NASA radiation codes NEQAIR [11] and HARA [30-31] as highlighted in Figure 12 where the computed and measure radiances for each range are plotted.

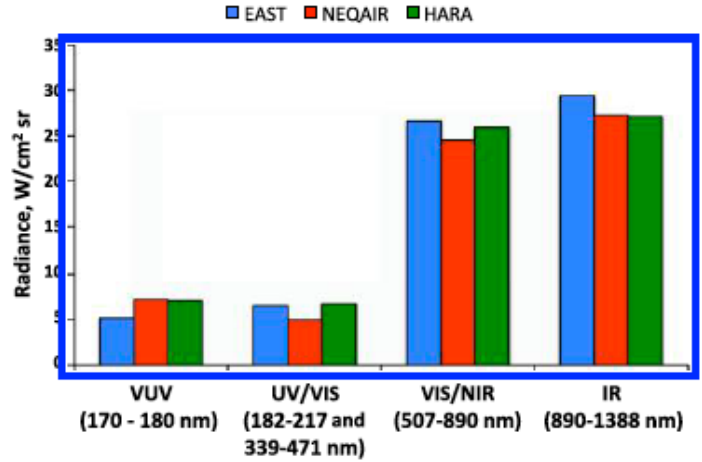

**Figure 12: Computed and measured radiance in EAST at 11 km/s (from [25])**

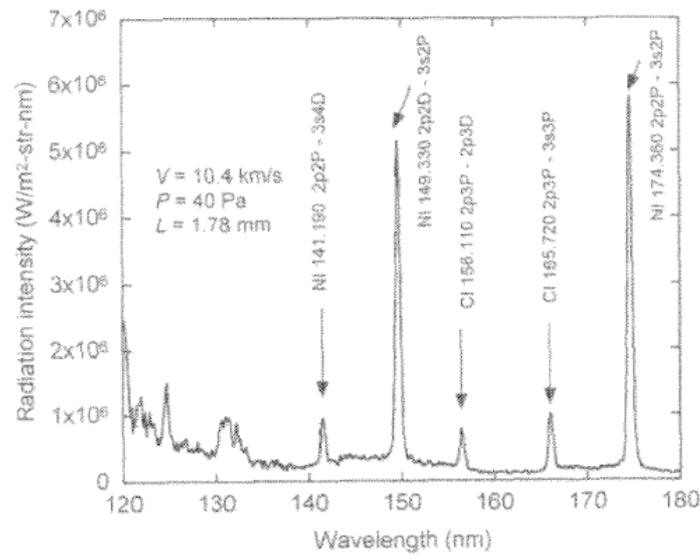

**Figure 13: Spectrum of the VUV region in pure nitrogen at 10.4 km/s and 40 Pa (from [22])**

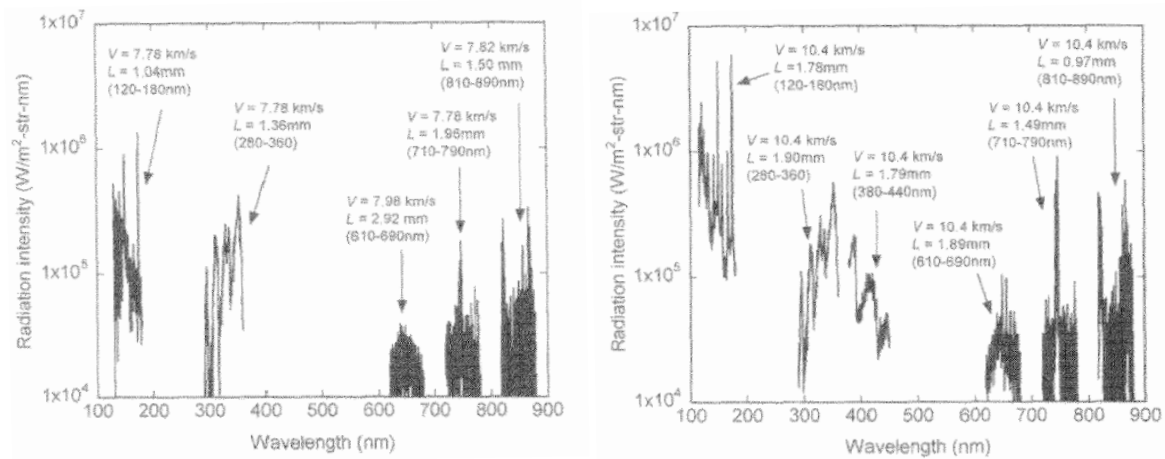

**Figure 14: Spectra obtained and 7.8 km/s (left) and 10.4 km/s (right) (from [22])**

Some of the tests carried out in EAST have also been analysed, in the VUV and IR ranges at JAXA by Lemal et al [32] for shock velocities of 10.54 and 11.17 km/s. The objective was to assess a collisional-radiative model for VUV and IR ranges. Good agreement was obtained for the IR and VUV intensity profiles, for both conditions after a tuning of the model for accounting heavy particle impact.

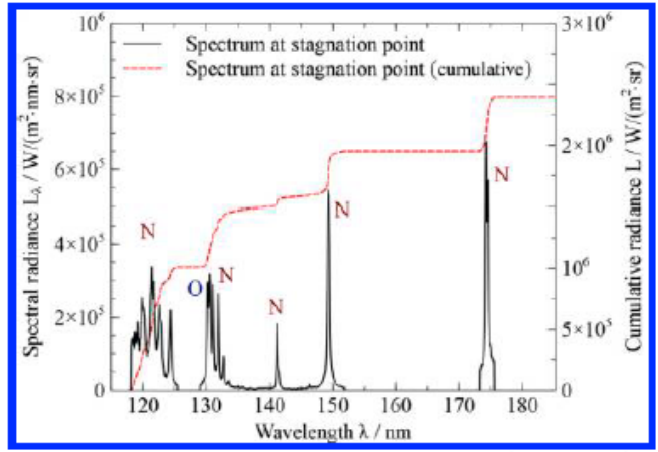

**Figure 15: Experimental VUV radiation in X2 for 12.2 km/s (from [34])**

|  | High power condition |
|---|---|
| Gas | Air |
| Freestream pressure | 1atm |
| Plate power | 62 kW |
| Maximum temperature 27.5 mm from exit | 6708 K |
| Specific enthalpy 27.5 mm from exit at $T_{max}$ | 21.8 MJ/kg |
| Species mole fractions 27.5 mm from exit at $T_{max}$: |  |
| O | 0.23 |
| N | 0.28 |
| $O_2$ | $1.1 \times 10^{-4}$ |
| $N_2$ | 0.47 |
| CN | $5.9 \times 10^{-6}$ |
| CO | $2.3 \times 10^{-4}$ |

**Table 2: Experimental conditions for plasma torch tests (from [37])**

To support the Hayabusa mission, test campaigns have been performed using the JAXA HVST shock-tube and pure nitrogen [22]. Emission intensity from VUV to NIR has been investigated for a pure nitrogen atmosphere and two experimental conditions 120 Pa and 7.8 km/s, and 40 Pa and 10.4 km/s. These two points are located along the Hayabusa re-entry trajectory. Results show that C and N atomic lines are predominant in the VUV region (see Figure 13) and no molecular bands are seen. The UV region is dominated by $N_2(2+)$ and $N_2^+(1-)$ systems, while atomic lines of N are intense in the visible. The results show that the spectrum of the VUV region becomes more intense when increasing the shock velocity as highlighted in Figure 14. The experimental data were numerically reconstructed and radiation analysis performed with SPRADIAN2 [33] and good agreement was obtained when the electronic temperature was higher than the vibrational one.

Super-orbital re-entry flow conditions [34] were investigated in the X2 expansion tunnel at the University of Queensland, corresponding to 12.2 km/s and a static pressure of 870 Pa. VUV measurements were performed for a wavelength range of 116-185 nm using a deuterium lamp. The corresponding spectrum is shown in Figure 15. Population densities were estimated by analysing the radiative transport in the flow, temperature from the measured spectra. The experiment was numerically reconstructed using CFD URANUS code [35], and PARADE line-by-line code [36]. The numerical results obtained are in a good agreement with the electron density and temperature derived from the measurements.

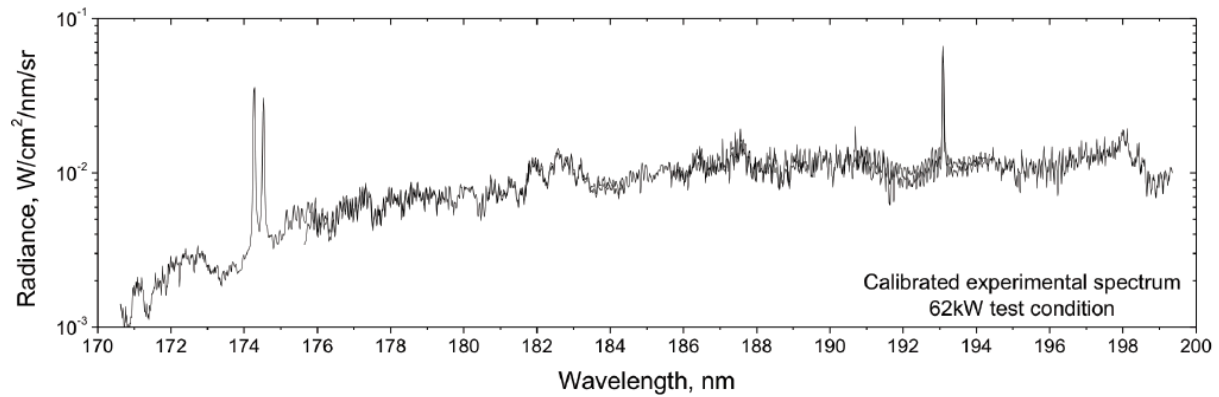

Figure 16: Measured calibrated spectrum (from [37])

## 2.2 Tests with air in plasma facilities

Several test campaigns have been carried out at Ecole Centrale de Paris [37-38], CNRS Orléans [39] and IRS [40-41] for performing VUV measurements in air. This development of the VUV measurement technique, in some European plasma facilities, is due, until now, to the lack of a high velocity shock facility in Europe with the capability of reproducing superorbital re-entry conditions.

Jacobs et al [37-38] have performed spectroscopy measurements in the VUV region for air plasma in an ICP plasma torch. Measurements were conducted in the range of 170-200 nm for equilibrium emission conditions at 6708 K and a specific enthalpy of 21.8 MJ/kg. The test conditions are summarized in Table 2. The measured calibrated spectrum is displayed in Figure 16. The measurements have been numerically reconstructed using the line-by-line radiation code SPECAIR [42] and little differences found between the experimental data and the numerical predictions as shown in Figure 17.

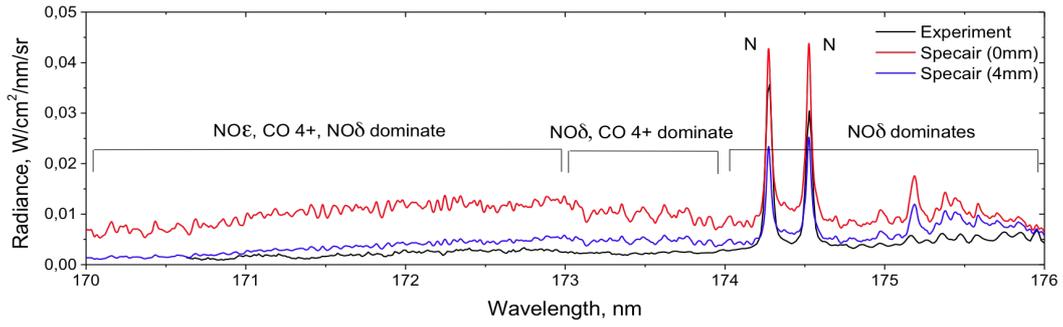

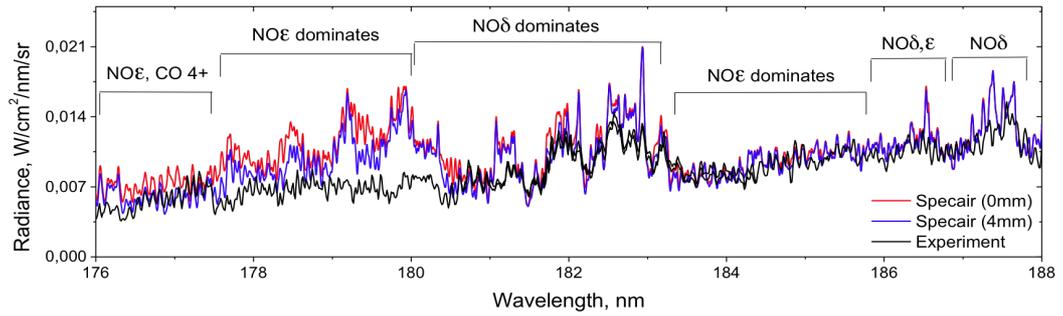

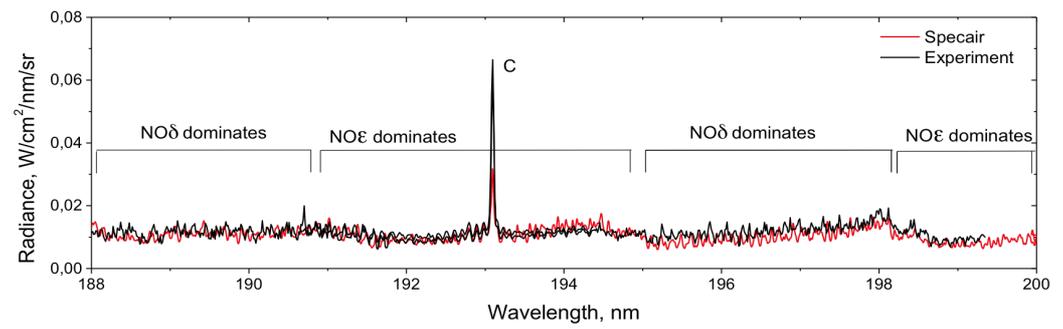

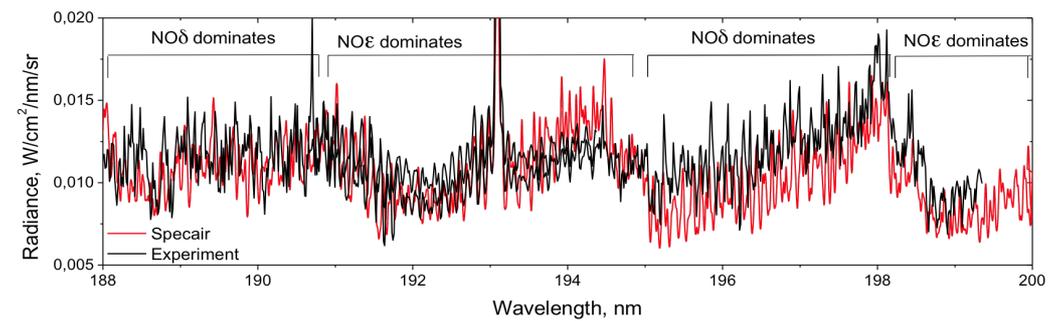

**Figure 17: Experimental data and SPECAIR calculations (from [37])**

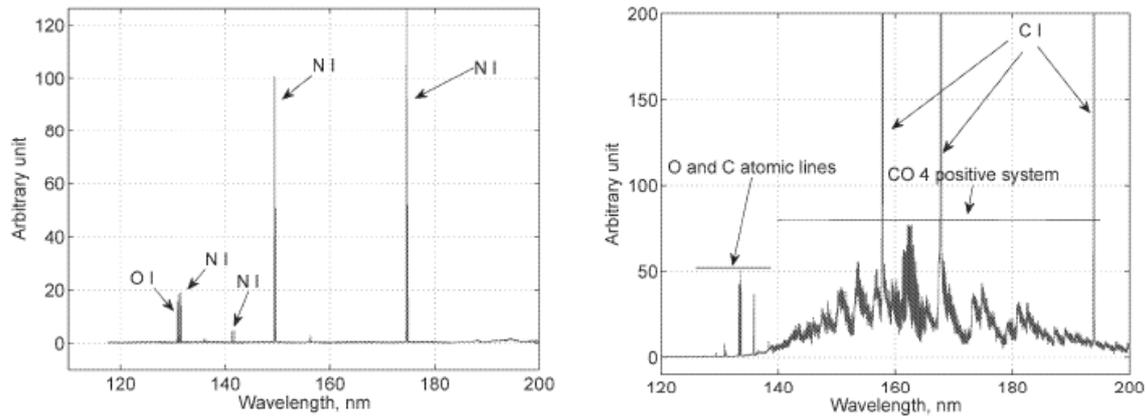

**Figure 18: VUV measurements for air plasmas (from [39])**

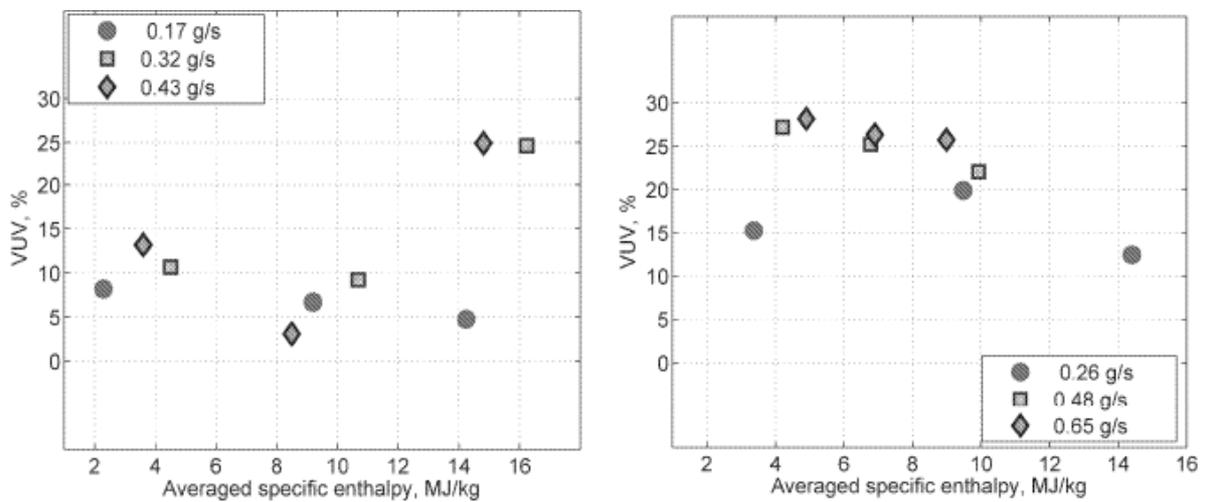

**Figure 19: Contribution of VUV range to the total radiation as function of specific enthalpy (left air, right Mars atmosphere) (from [39])**

Air (80 % $N_2$ and 20 $O_2$) plasma flows have been investigated at CNRS Orléans in the PHEDRA low-density plasma wind-tunnel at ICARE (former SR5 facility at CNRS Orléans) by Lago [39]. The recent test campaign performed by Lago [39] focused on VUV radiation for both Earth and Mars entry conditions. Spectral emission was measured from 110 to 900 nm in low-pressure flows: pressure chamber was in between 2 and 3.9 Pa for air. VUV spectrum is shown in Figure 18 but the exact flow conditions for which it has been obtained are not clearly defined in [39]. For air, VUV radiation presents only atomic lines for N and O since no molecular system was identified, as highlighted in Figure 18. VUV contribution to the total radiation was evaluated for the different flow conditions. Results are summarized in Figure 19, on the left for air and on the right for a Mars atmosphere. For low enthalpies (2-4.5 MJ/kg), the VUV contribution ranges from 8 to 14 %, for medium enthalpies (8.5 – 11 MJ/kg a contribution in between 2 and 10 % is observed, while at high enthalpy (15-16 MJ/kg) VUV radiation reaches 25 % of the total radiation which is close to the level observed in some shock-tube experiments for super-orbital re-entry conditions [**Erreur ! Signet non défini.**].

| Static pressure (Pa) | 1660 |
|---|---|
| Heat-flux (MW/m$^2$) | 4.4 |
| Specific enthalpy (MJ/kg) | 68.4 |
| Temperature (K) | 11780 |
| Velocity (m/s) | 3350 |

**Table 3: PWK1 test conditions 20 mm in front of the probe [40-41]**

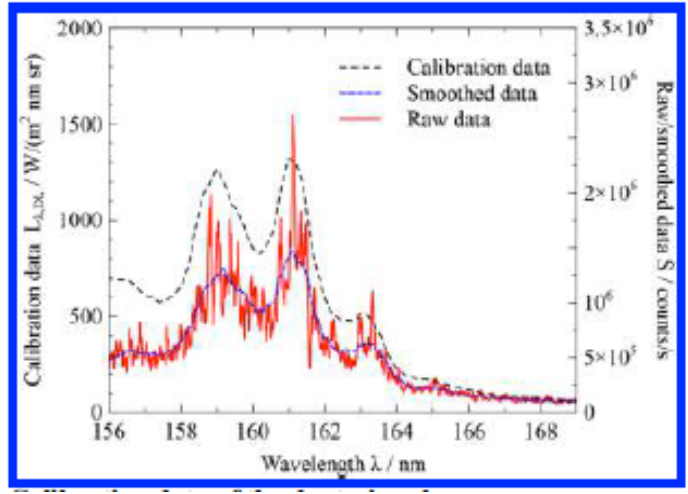

**Figure 20: Calibration results for the VUV range using a deuterium lamp (from [40])**

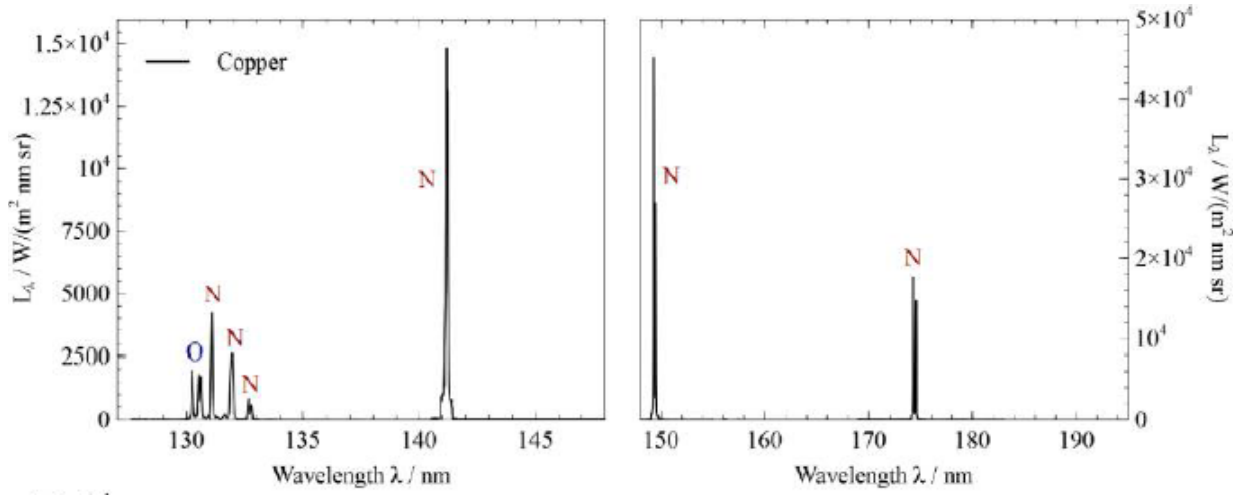

**Figure 21: Measured VUV radiation at stagnation point for a copper sample (from [41])**

The test conditions reported in Table 3 has been investigated in the plasma wind-tunnel PWK1 at IRS [40-41]. They correspond to the point at 78.8 km of altitude and 11.7 km/s of the Hayabusa re-entry trajectory. Radiation measurements were carried out from 120 nm up to 960 nm using a cooled copper probe in [40], and a carbon phenolic sample in [41]. Calibration results with raw data, calibrated data, and smooth data, are shown in Figure 20. The contribution of the VUV range to the radiative heating measured at the stagnation point for the conditions reported in Table 3 was 235 kW/m². VUV radiation is dominated by nitrogen and oxygen atomic lines. This is highlighted in Figures 21 and 22, displaying the measured VUV radiation as function of the wavelength, for a copper, and a carbon phenolic probe respectively. It has to be noted that the presence of the ablative sample has a strong influence on the VUV radiation that is reduced. This is due to the reduction of atomic nitrogen through the decomposition of the carbon phenolic and the blocking effect [43] generated by the pyrolysis gases.

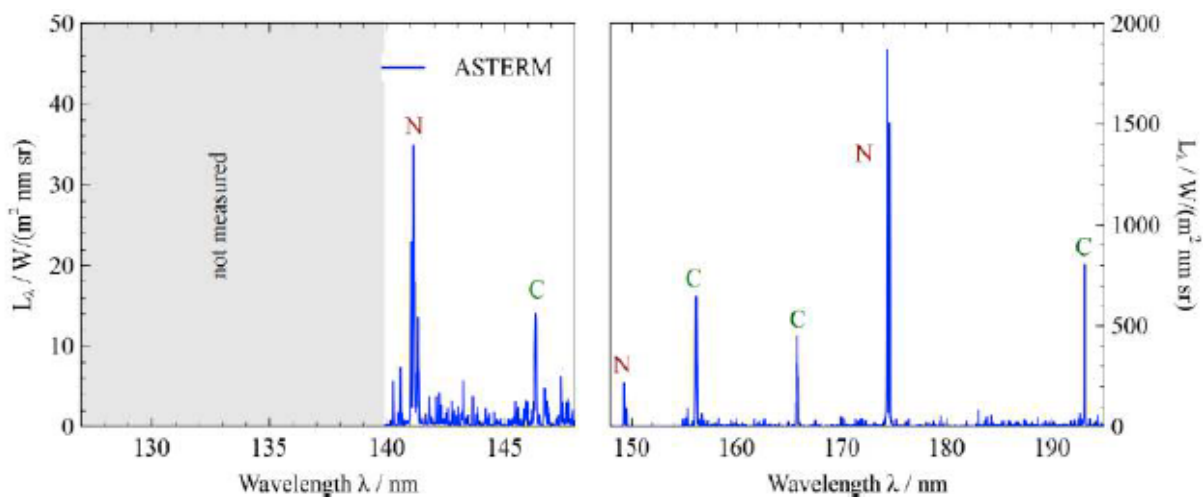

**Figure 22: Measured VUV radiation for a carbon phenolic (ASTERM) probe (from [41])**

## 3 Discussion

A survey of the available experimental datasets for VUV radiation has been carried out. Due to the limited number of studies performed for atmospheric entry, the effort has been extended to the other domain of plasma physics. However, the most relevant data obtained for VUV radiation are those obtained in the frame of studied undertaken to support atmospheric entry investigations. Several datasets covering the VUV range for Earth superorbital entry have been identified; some have been sufficiently detailed for possible reproduction in other ground tests. Some of the corresponding test campaigns were performed in shock-tubes, the other using plasma torches. However, in the frame of ESTHER development, the tests performed in shock facilities will be the easier to duplicate. This will allow future comparisons with ESTHER results as soon as they are available.

In the USA, EAST and LENS-XX have already provided valuable data for Earth high-speed re-entry. For EAST a large number of shots has been performed for Lunar return conditions [44] and data were collected from UV to near infrared. Similar shock-velocities (around 9-10 km/s) have been simulated with LENS-XX [45], at very low pressure (26.6 Pa), and shock emission was investigated for a wavelength range in between 200 nm to 1000 nm, from UV to infrared bands, but VUV measurements are missing from the collected data.

Other VUV test campaigns have been conducted in the JAXA HVST shock-tube. Results have been obtained for both Earth high-speed entry and Mars entry conditions. In Europe, due to the lack of an adequate facility test campaigns have been led in plasma torches in France at CORIA, ICARE, and Ecole Centrale de Paris, and in Germany at IRS. If the corresponding datasets will be difficult to duplicate in a high-velocity shock tube, most of them have been obtained for subsonic conditions, the data obtained will be useful for upgrading the existing radiation databases. Moreover, during long years, these facilities have allowed maintaining in Europe, experimentalists and expertise in radiation measurements, which will be an asset for the use of ESTHER. Additionally the use of such smaller facilities is very relevant for developing new measurement techniques, to be used or adapted, later on, in ESTHER.

Concerning the comparisons between the future data obtained with ESTHER and the existing ones. So far, similar exercise has already been performed with some shock velocity facilities. X2 experimental data have already been compared against those obtained in EAST and HVST for Mars entry conditions [46], and against EAST results for Earth high-speed entries [47]. However, if these former comparisons did not account for VUV radiation, this makes these two last facilities among the most attractive for a similar exercise with ESTHER.

## 4 Conclusions

An extensive screening of VUV measurements have been carried out for identifying experimental datasets valuable for possible cross-checks with future ESTHER measurements. The review has been extended to other domains than atmospheric entry in order to potentially enlarge the number of datasets and also to identifying other fields of interest for the shock-tube activity. However, the most attractive data remain those obtained for supporting space exploration missions.

This has allowed the identification of the datasets for future comparisons with ESTHER. Among the similar facilities, VUV data obtained with EAST and HVST are very interesting since VUV data obtained with these two shock-tubes have already been compared. In addition, potential comparisons could be also performed with the datasets obtained using X2, but for wavelengths outside the VUV range.

All these elements present certain usefulness for the future ESTHER calibration. The data presenting the major interest for potential cross-checking are those obtained with HVST, EAST, and to a lesser extend X2.

**Acknowledgment**

The research leading to these results has been partially supported by Fluid Gravity Engineering and Instituto Superior Tecnico of Lisbon under European Space Agency contract 23086.